\documentclass{article}

 \usepackage[final]{nips_2017}

\usepackage[utf8]{inputenc} 
\usepackage[T1]{fontenc}    
\usepackage{hyperref}       
\usepackage{url}            
\usepackage{booktabs}       
\usepackage{amsfonts}       
\usepackage{nicefrac}       
\usepackage{microtype}      
\usepackage{graphicx}
\usepackage{subfig}
\usepackage{xspace}

\title{Scaling GRPC Tensorflow on up to 512 nodes of Cori Supercomputer}

\author{
  Amrita Mathuriya \\
  Intel Corporation \\
  \texttt{amrita.mathuriya@intel.com}
  \And 
  Thorsten Kurth \\
  Lawrence Berkeley \\ National Laboratory 
  \And
  Vivek Rane \\
  Intel Corporation 
  \And
  Mustafa Mustafa \\
  Lawrence Berkeley \\ National Laboratory 
  \And
  Lei Shao \\
  Intel Corporation 
  \And
  Debbie Bard \\
  Lawrence Berkeley \\ National Laboratory 
  \And
  Prabhat \\
  Lawrence Berkeley \\ National Laboratory 
  \And
  Victor W Lee \\
  Intel Corporation 
}

\newcommand{\registered}{\textsuperscript{\textregistered}\xspace}
\newcommand{\trademark}{\texttrademark\xspace}

\newcommand{\CompilerDisclaimer}{
Optimization Notice: Intel's compilers may or may not optimize to the same
degree for non-Intel microprocessors for optimizations that are not unique to
Intel microprocessors. These optimizations include SSE2, SSE3, and SSSE3
instruction sets and other optimizations. Intel does not guarantee the
availability, functionality, or effectiveness of any optimization on
microprocessors not manufactured by Intel. Microprocessor-dependent
optimizations in this product are intended for use with Intel microprocessors.
Certain optimizations not specific to Intel microarchitecture are reserved for
Intel microprocessors. Please refer to the applicable product User and
Reference Guides for more information regarding the specific instruction sets
covered by this notice.
}

\newcommand{\BenchmarkDisclaimer}{
Software and workloads used in performance tests may have been optimized for
performance only on Intel microprocessors.  Performance tests, such as SYSmark
and MobileMark, are measured using specific computer systems, components,
software, operations and functions. Any change to any of those factors may
cause the results to vary. You should consult other information and performance
tests to assist you in fully evaluating your contemplated purchases, including
the performance of that product when combined with other products.  For more
complete information visit http://www.intel.com/performance.
\\
Intel, Xeon, and Intel Xeon Phi are trademarks of Intel Corporation in the U.S. and/or other countries.
}

\begin{document}
\maketitle


\section{Introduction}\label{sec:intro}
We explore scaling of the standard distributed Tensorflow~\cite{TFPub} with GRPC primitives on up to 512 Intel\registered Xeon Phi\trademark (KNL) nodes of Cori supercomputer~\cite{Cori} with synchronous stochastic gradient
descent (SGD), and identify causes of scaling inefficiency at higher node counts. 
To our knowledge, this is the first exploration of distributed GRPC Tensorflow's scalability on a HPC supercomputer at such large scale with synchronous SGD.
We studied scaling of two convolution neural networks - ResNet-50~\cite{resnet50}, a state-of-the-art deep network for classification with roughly 25.5 million parameters, and HEP-CNN~\cite{hepGithub}~\cite{cite15PF}, a shallow topology with less than 1 million parameters for common scientific usages.
For ResNet-50, we achieve >80\% scaling efficiency on up to 128 workers, using 32 parameter servers (PS tasks) with a steep decline down to ~23\% for 512 workers using 64 PS tasks. 
Our analysis of the efficiency drop points to low network bandwidth utilization due to combined effect of three factors. (a) Heterogeneous distributed parallelization algorithm 
which uses PS tasks as centralized servers for gradient averaging is suboptimal for utilizing interconnect bandwidth. (b) Load imbalance among PS tasks hinders their efficient scaling. (c) Underlying communication primitive GRPC is currently inefficient on Cori's high-speed interconnect.
The HEP-CNN demands less interconnect bandwidth, and shows >80\% weak scaling efficiency for up to 256 nodes with only 1 PS task.
Our findings are applicable to other deep learning networks. Big networks with millions of parameters stumble upon the issues discussed here. Shallower networks like HEP-CNN with relatively lower number of parameters can efficiently enjoy weak scaling even with a single parameter server.


\section{Configuration}
We used Intel\registered Xeon Phi\trademark ``Knights Landing'' (KNL) processors 
from Cori supercomputer (Phase II) with GPFS file system. 
We use dummy data and do not perform disk I/O during the runs.
The interconnect used here is 
Cray ``Aries'' high speed inter-node network with Dragonfly 
topology with 45.0 TB/s global peak bisection bandwidth for 9,688 KNL compute nodes (Phase II).
The single node thread and affinity settings used are: 
  KMP\_BLOCKTIME = 0, KMP\_SETTINGS = 0, 
  KMP\_AFFINITY = ``granularity=fine,noverbose,compact,1,0'' and 
  OMP\_NUM\_THREADS = 66
  Num\_inter\_threads = 66 and Num\_intra\_threads = 3.
  For ResNet-50 experiments, we use standard tf\_cnn\_benchmarking scripts~\cite{TFBM}.
We do minimal changes to the script to enable distributed Tensorflow run on Cori with the Slurm schedular 
and running on CPU only mode. These include defining local\_parameter\_device and 
device parameters as `cpu' and setting force\_gpu\_compatible and use\_nccl to false.
For the HEP-CNN benchmark~\cite{hepGithub}, we update the code 
to use the tf.train.Supervisor API.
The Tensorflow~\cite{TFCode} code was compiled from sources with MKL~\cite{mkl} optimizations on August 29. 
Tensorflow version is 1.3 and compiler used is  gcc/6.3.0.

\section{Experiments}
Our experiments\footnote{\BenchmarkDisclaimer}
do not aim to research scaling algorithms, but rather to assess 
the scaling issues of GRPC distributed TensorFlow in 
production on Cori at large scale as experienced by its users. 
We (a) study weak scaling (fixed minibatch size per worker) with synchronous SGD algorithm; (b) study the distributed parameter update algorithm with PS tasks and workers using GRPC communication primitives; (c) choose two networks with different communication characteristics; (d) choose a relatively large batch-size of 128 images per worker to keep the fraction of communication time low at low worker counts; (e) use dummy data to avoid any potential I/O bottlenecks.
We are using 
Tensorflow
compiled with MKL~\cite{mkl} 
which is optimized\footnote{\CompilerDisclaimer} 
for Intel\registered CPUs.
The two benchmarks chosen ResNet-50 and HEP-CNN perform classification tasks for images. 


\begin{figure*}[!ht]
\includegraphics[width=\textwidth]{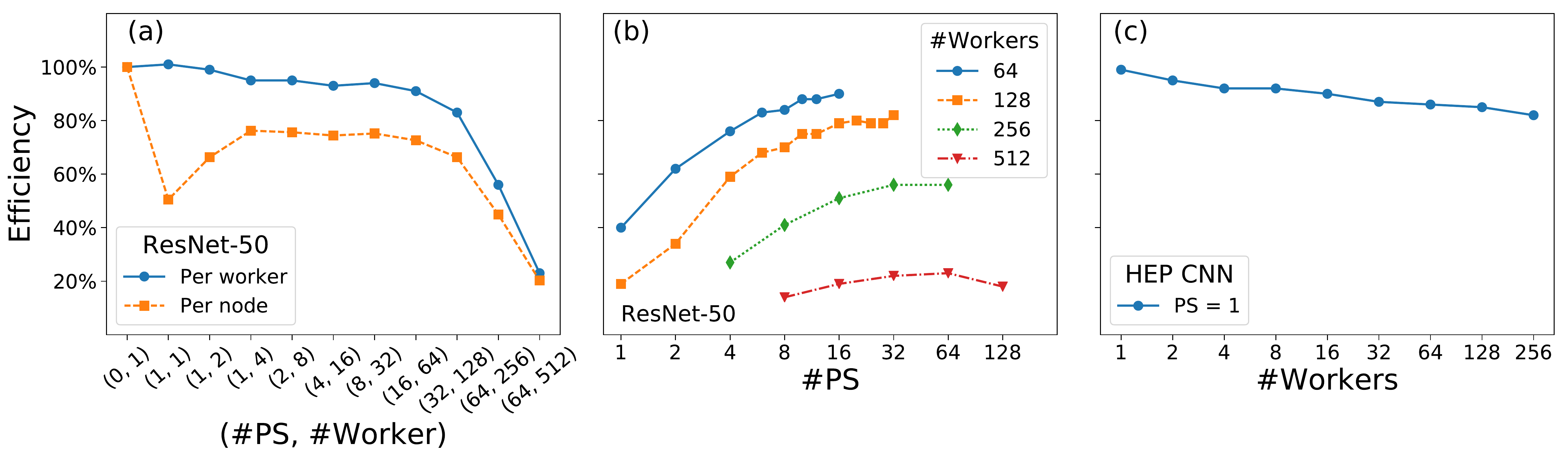}
\caption{\label{fig:exp} Scaling efficiency of GRPC distributed Tensorflow on Cori KNL nodes with 128 batch size per worker. Single KNL performance is chosen as the baseline
All PS tasks and workers execute on different nodes. Both benchmarks use (224, 224, 3) size and NCHW format for input images. (a) Efficiency for ResNet-50 vs. number of workers. (b) Efficiency for ResNet-50 vs. number of PS tasks (c) Efficiency of HEP-CNN vs. number of workers.} 
\end{figure*}
 
Figure~\ref{fig:exp}~(a) shows the efficiency of scaling ResNet-50 on up to 512 workers. The number of PS tasks are chosen for highest per worker efficiency, shown in Fig.~\ref{fig:exp}~(b). 
We find that the aggregate performance of all workers increases with an increase in the number of PS tasks, and then flattens or decreases. Figure~\ref{fig:exp}~(a) shows that higher than 80\% per worker efficiency can be obtained with up to 128 workers supported by 32 PS nodes. However, the efficiency drops to a maximum of 56\% for 256 workers and 23\% for 512 workers; and it does not improve with higher number of PS tasks. 
A similar trend is observed using 2012 ILSVRC competition dataset~\cite{ilsvrc2012}. For the HEP-CNN, 1 PS task is capable of supporting up to 256 worker nodes with >80\%  efficiency, Fig.~\ref{fig:exp}~(c). 

\section{Analysis}
The two neural networks show different scaling characteristics.
ResNet-50 is a 50 layer deep network and contains millions of parameters, whereas HEP-CNN is a light weight network with a total of 6 layers and roughly 593K parameters.
Our analysis suggests that at large node counts, scaling of ResNet-50 network becomes bottlenecked by the achievable interconnect bandwidth. 
This is due to various inefficiencies in the current GRPC based distributed parameter update algorithm  which result in suboptimal use of the interconnect bandwidth.
The parameter update algorithm distributes trainable variables of each layer to centralized servers.  Each server is responsible for combining updates of a set of variables from all workers.   This scheme has two problems. (1) Due to the synchronous nature of the training process, this can introduce hotspots in the interconnect when all workers send updates to the same server at roughly the same time. The amount of traffic to each server increases linearly with the number of workers.  At some point, the interconnect bandwidth of PS tasks becomes the bottleneck of scaling. (2) Load imbalance among the PS tasks, as the number of PS tasks is limited to the number of disjoint parameter sets in the current benchmarks.




One can reduce the interconnect bandwidth bottleneck by allocating more nodes to PS tasks.
In our experiments, we need to dedicate as many as 1/4 additional nodes to PS tasks to achieve >80\% per worker scaling efficiency for ResNet-50 (Fig.~\ref{fig:exp}~(b)).
However, this reduces per-node efficiency as shown in Fig.~\ref{fig:exp}~(a).
HEP-CNN puts less pressure on the interconnect bandwidth and we do not see significant performance improvement with the increase in PS tasks. 
Also, load imbalance among PS tasks does not allow efficient weak scaling to continue beyond 128 workers for ResNet-50 network, even with 32 or higher PS tasks. 
In ResNet-50, 99\% of the ~25.5M parameters are contained in 54 two or higher dimensional tensors. Tensors are assigned to PS tasks using a greedy load balancing strategy based on their sizes. Each tensor variable is assigned to a single PS task only. Given this type of allocation strategy, scaling PS tasks beyond 54 results in heavy load imbalance among the PS nodes. Our experiments clearly show (Fig.~\ref{fig:exp}~b) that the gain from increasing PS tasks from 32 to 64 results in insignificant performance improvement. 

The current GRPC protocol limits the maximum amount of bandwidth that can be utilized by each node. 
Our crude estimates show roughly 5-6x gap in the communication time for ResNet50 for 1 PS and 16 workers compared to the peak achievable.
Our conjecture is that improving the communication protocol (such as using MPI) would improve overall scaling.

\section{Outlook}
Our analysis indicates suboptimal use of high-speed interconnect bandwidth by the current distributed algorithm and implementation of Tensorflow in production at Cori, which uses the GPRC protocol. 
There are more optimal ways of implementing all-reduce operation, such as tree-reduction or the ring method~\cite{mpichCollective}~\cite{BiaduRing} which has lower theoretical complexity than a linear dependence on the number of nodes.
In future work, we plan to explore those and MPI communication primitives.  

\section*{Acknowledgment}
We thank Jeongnim Kim, Bhavani Subramanian, Mahmoud Abuzaina, Lawrence Meadows, Elmoustapha Ould-ahmed-vall 
and AG Ramesh for their help in discussions and set up.  
This work is a part of of the Intel/NERSC Big Data Center collaboration. 

\bibliographystyle{IEEEtran}
\bibliography{sample}

\end{document}